# Phase Dependent Thermopower in Andreev Interferometers

Jonghwa Eom, Chen-Jung Chien, and Venkat Chandrasekhar

*Department of Physics and Astronomy, Northwestern University, 2145 Sheridan Road, Evanston, IL 60208*

(November 6, 1997)

We report measurements of the thermopower $S$ of mesoscopic Andreev interferometers, which are hybrid loops with one arm fabricated from a superconductor (Al), and one arm from a normal metal (Au). $S$ depends on the phase of electrons in the interferometer, oscillating as a function of magnetic flux with a period of one flux quantum $\Phi_0 = h/2e$. The magnitude of $S$ increases as the temperature $T$ is lowered, reaching a maximum at $T \sim 0.14$ K, and decreases at lower temperatures. The symmetry of $S$ oscillations with respect to magnetic flux depends on the topology of the sample.

74.50.+r, 73.23.-b, 74.25.Fy

Recently, many theoretical papers [1] have addressed the question of transport in hybrid mesoscopic structures in which some part of the device is a superconductor, and the remainder is either metallic or formed from a semiconductor heterostructure device. A number of experiments [2–4] in the past few years have demonstrated the variety of phenomena that can be observed in these hybrid devices. The fundamental physics that forms the basis for many of these effects arises from the interaction of electrons in the metal or semiconductor with Cooper pairs in the superconductor at the normal metal/superconducting (NS) interface. At temperatures below the transition temperature $T_c$ of the superconductor, the interaction takes place primarily by the process of Andreev reflection [5]. An electron in the normal metal approaching the interface with energy less than the gap of the superconductor cannot be transmitted, but is phase coherently reflected as a hole with the simultaneous generation of a Cooper pair in the superconductor. Many recent experiments [3] on so called Andreev interferometers (which are mesoscopic hybrid loops in which one arm is formed from superconductor and one arm from a normal metal) have demonstrated beautifully the phase coherent nature of Andreev reflection.

Most experiments on mesoscopic NS devices to date have focused on measuring the electrical conductance. However, unusual effects have also been predicted for other properties such as the thermopower and thermal conductivity [6]. In this Letter, we report the first measurements of the thermopower of mesoscopic Andreev interferometers. We find the thermopower depends on the phase of electrons in the interferometer, oscillating as a function of magnetic field $B$ with a fundamental period corresponding to one superconducting flux quantum $\Phi_0 = h/2e$ through the loop. The symmetry of the thermopower oscillations with respect to magnetic field depends on the topology of the sample, and is not necessarily the same as that of the magnetoresistance oscillations. The magnitude of the thermopower, which is on the order of a few $\mu V/K$, shows a non-monotonic dependence on temperature, with a maximum around a temperature $T \sim 0.14$ K, but this non-monotonic behavior is not associated with the reentrance effect recently observed in the electrical transport properties of NS devices [4].

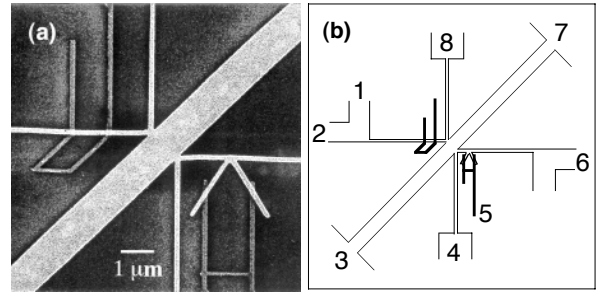

**Fig. 1.** (a) Scanning electron micrograph of a typical sample. (b) Schematic of the sample structure. Fat solid lines represent Al wires, while thin solid lines outline the Au film. The two hybrid loops are labeled as 'parallelogram' (left) and 'house' (right). The contacts are labeled with Arabic numerals.

The design of the samples for this experiment was strongly influenced by our experience in measuring thermoelectric effects in mesoscopic spin-glasses [7]. Figure 1 shows a schematic of our sample, along with a scanning electron micrograph of one of the samples. The sample consists essentially of a wide Au strip ($\sim 25$ $\mu m$ long, $\sim 1$ $\mu m$ wide) with four Au probes (each $\sim 7.2$ $\mu m$ long, $\sim 0.1$ $\mu m$ wide) emanating from its center. On two of these probes (the horizontal probes), a second layer of lithography is used to define two types of Andreev interferometers, as shown in Fig. 1. We shall denote the Andreev interferometer to the right of the wide Au strip the 'house,' and the one to the left of the strip the 'parallelogram.' The wide metallic strip acts as a heater, raising the temperature $T_e$ of the electrons above the bath temperature $T$ when a dc current $I$ is passed through it via leads 7 and 3 (see Fig. 1(b)). The electrons are cooled by phonons in the wide strip through the electron-phonon interaction, and also by electronic thermal conduction to the large contacts of the strip, which are at the bath temperature $T$. Electronic cooling is most effective near the contacts, and leads to a non-uniform electron tempera-ture profile $T_e(x)$ as a function of the distance $x$ along the strip. Taking into account all these contributions, $T_e$ is given by the solution of Nagaev's equation [8]

$$\frac{\pi^2}{3}L^2 k_B^2 \frac{d}{dx}(T_e \frac{dT_e}{dx}) = -(eIR)^2 + \frac{\gamma k_B^2}{\Theta_D^3}(T_e^5 - T^5) \quad (1)$$

where $L$ is the total length of the wire, $R$ the resistance, $\gamma = 24\zeta(5)\alpha_{ph}L^2 k_B \Theta_D/\hbar D$, $\alpha_{ph}$ the dimensionless electron-phonon coupling constant, $\Theta_D$ the Debye temperature, $D = v_F l_{el}/3$ the diffusion constant, $v_F$ the Fermi velocity, $l_{el}$ the elastic mean free path, and $\zeta$ the Riemann zeta function. Consider now the thermoelectric voltage $V_{th}$ developed between two probes on the same side of the wide strip, one of which is pure Au, and one which contains an Andreev interferometer (e.g., contacts 1 and 8 in Fig. 1(b)). Assuming that the large contacts at the end of each probe are at the bath temperature $T$, $V_{th}$ developed between the contacts is given by

$$V_{th} = \int_T^{T_m} S_A \, dT_e + \int_{T_m}^T S_N \, dT_e \approx \int_T^{T_m} S_A(T_e) \, dT_e \quad (2)$$

where $T_m$ is the temperature of the electrons in the middle of the metallic strip at the junction of the two probes, $S_A$ the thermopower of the probe containing the Andreev interferometer, and we have assumed that the thermopower $S_N$ of the normal probe is negligible at low temperatures [9]. In our measurements, we superpose a small low frequency ac signal on top of the dc current $I$ to increase our sensitivity. The differential resistance arising from $V_{th}$ in Eq. (2) is then given by

$$R_{th} = \frac{dV_{th}}{dI} = S_A \frac{dT_m}{dI} \quad (3)$$

since $T$ is independent of the current $I$. It is important to note that $R_{th}$ is a purely antisymmetric function of the current $I$ because $T_m$ is independent of the direction of the dc current. The antisymmetric component of the total differential resistance does not contain contributions from the transverse or longitudinal resistance of the wide strip, because these would contribute only to the symmetric component. Hence, by measuring the differential resistance in this configuration with a small but finite dc current, one directly obtains information about the thermopower of the voltage probes.

Measurements on two samples (A and B) are reported here. The samples were patterned by conventional two-level electron-beam lithography. After the first level of lithography, the metallic layer was deposited in one evaporation of 99.999% pure Au onto an oxidized Si substrate. Following liftoff and a second level of lithography, 99.999% pure Al was deposited to form the superconducting layer. In order to ensure a good NS interface, the samples were etched with Ar$^+$ ions *in situ* prior to the Al evaporation. The area of the parallelogram was $\sim$ 1.3 $\mu m^2$ and the area of the house $\sim 4.1$ $\mu m^2$. The thicknesses of the metal films (Au/Al) for the samples were Au : 55 $nm$, Al : 80 $nm$ for sample A, and Au : 55 $nm$, Al : 60 $nm$ for sample B respectively. Narrow Au and Al wires of the same nominal linewidth as the probes, as well as separate Au/Al interfaces were fabricated simultaneously on the same chips in order to determine relevant film parameters for each sample. The resistance of the control NS interface was less than $\sim 0.4$ $\Omega$ for both samples. From measurements on the control samples, other relevant film parameters are as follows: Sample A, sheet resistance $R_\square$ at 4.2 K $\approx 0.24$ $\Omega$, coherence length in the normal metal $L_T = \sqrt{\hbar D/k_B T} \approx 0.48$ $\mu m$ at $T = 1$ K, and electron phase coherence length $L_\varphi \approx 6.7$ $\mu m$ at 351 mK; Sample B, $R_\square \approx 0.35$ $\Omega$, $L_T \approx 0.40$ $\mu m$ at $T = 1$ K, and $L_\varphi \approx 4.6$ $\mu m$ at 38 mK. Measurements were performed in a $^3$He cryostat for the parallelogram structure of sample A, and in a dilution refrigerator for the house structure of sample B. In addition to the thermopower measurements, the magnetoresistance of the Andreev interferometers was measured using conventional ac bridge techniques.

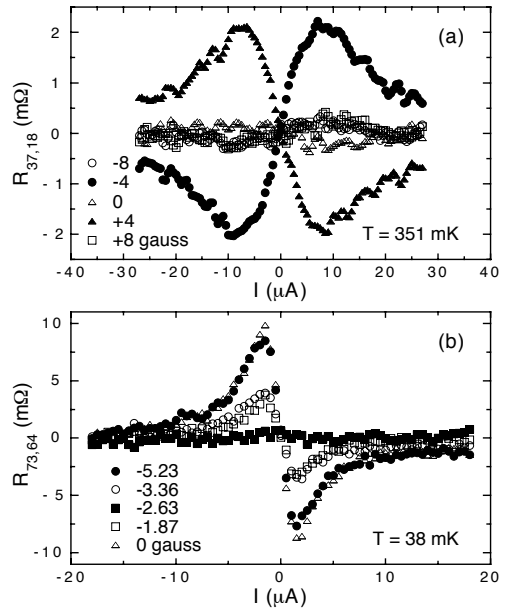

**Fig. 2.** (a) Differential resistance $R_{37,18}$ of the parallelogram structure of sample A as a function of dc bias current $I$ at a magnetic field corresponding to a flux of approximately $-\Phi_0/2$, $-\Phi_0/4$, 0 $\Phi_0$, $\Phi_0/4$, and $\Phi_0/2$ through the loop. (b) Differential resistance $R_{73,64}$ of the house structure of sample B as a function of $I$ at a magnetic field corresponding to a flux of approximately $-\Phi_0$, $-0.66\Phi_0$, $-0.5\Phi_0$, $-0.38\Phi_0$, and 0 $\Phi_0$ through the loop.

Figure 2(a) shows the differential resistance $R_{37,18}$ of the parallelogram structure of sample A at five different values of magnetic field at $T = 351$ mK. (We shall use the standard notation $R_{ij,kl}$, where the indices $i$ and $j$ denote the current leads, and the indices $k$ and $l$ the voltage leads.) The magnetic field corresponding to $\Phi_0$ through the loop is 15.9 gauss. At zero field, the curve is essentially flat, showing that the thermopower of the parallelogram interferometer is comparable to that of the pure Au wire. At a field corresponding to a flux $\Phi_0/4$ ($\sim$ 4.0 gauss) through the loop, $R_{37,18}$ has its maximum amplitude, and is clearly antisymmetric in $I$ to within our noise. As the field is increased further to $\Phi_0/2$ ($\sim$ 8.0 gauss), $R_{37,18}$ is again essentially independent of dc current. If the field is increased in the opposite direction, similar behavior is seen, except that the sign of $R_{37,18}$ is switched. Figure 2(b) shows similar data taken at 38 mK for the house structure of sample B, for which $\Phi_0$ is 5.0 gauss. In contrast to the parallelogram structure, the house structure shows its maximum amplitude at zero field [10]. As the field is decreased to a flux $-\Phi_0/2$ ($\sim$ -2.6 gauss) through the loop, $R_{73,64}$ becomes independent of the dc current $I$, indicating that the thermopower at this field is essentially zero. As the field is decreased still further to $-\Phi_0$, $R_{73,64}$ retraces the zero field curve. Hence, although the antisymmetric differential resistance oscillates periodically with magnetic field for both the house and parallelogram structures, the phase of the oscillations for the two structures is different.

The difference in phase can be seen more clearly if we fix the dc current at a finite value and measure the differential resistance as a function of field [11]. Figure 3(a) shows these data for the parallelogram of sample A at a fixed dc current of 8.6 $\mu$A and $T = 351$ mK. For all these measurements, the zero of magnetic field was determined by simultaneous weak localization measurements on the Au control sample. As expected, $R_{37,18}$ oscillates periodically as a function of field, but is antisymmetric in magnetic field. For comparison, we also show the magnetoresistance of the same interferometer which is clearly symmetric in magnetic field. Figure 3(b) shows similar data for the house structure of sample B at a fixed dc current of 2.1 $\mu$A and $T = 38$ mK. For this sample, the differential resistance $R_{73,64}$ and the magnetoresistance are symmetric in field.

The symmetry of the oscillations can be understood if we take into account the topology of the samples [6]. Consider first the house structure. If $\phi_1$ is the phase of the superconductor at one NS interface induced by the magnetic field, and $\phi_2$ the phase at the other NS interface, the interference of Andreev reflected quasiparticles in the normal metal is determined by the phase difference $\Delta\phi = \phi_1 - \phi_2$. For this structure, reversing the magnetic field, i.e., making the transformation $\Delta\phi \to -\Delta\phi$, is equivalent to performing a reflection of the NS loop alone through its mirror-axis. Since this reflection does not affect the temperature gradient across the NS loop, the corresponding $V_{th}$ does not change, and hence one expects the thermopower to be symmetric with respect to magnetic field. For the parallelogram structure, on the other hand, the reflection changes the temperature gradient across the NS loop, and the corresponding $V_{th}$ changes its sign. Reversing $V_{th}$ is equivalent to a change in sign of the thermopower, and hence the thermopower is antisymmetric in the magnetic field. We note that this simple explanation is based on the assumption that the NS interfaces involved in this transformation are identical [6], which is not necessarily true in real samples. A conclusive explanation for the symmetry of the thermopower oscillation is yet to be made.

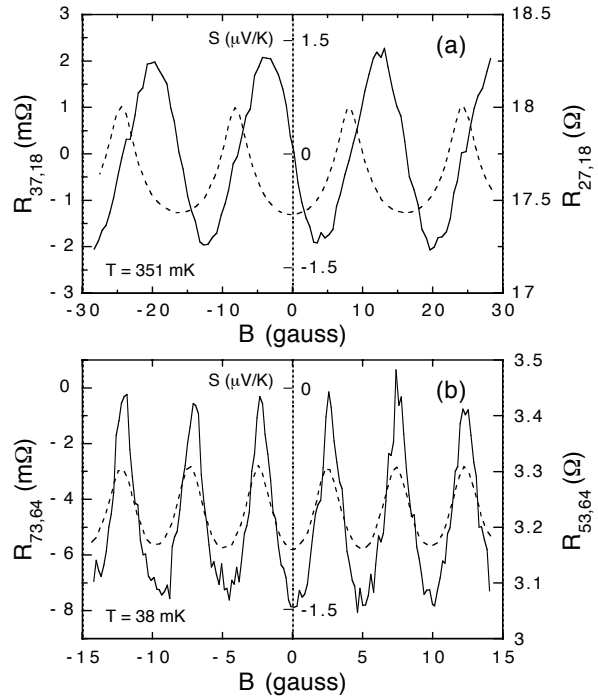

**Fig. 3.** (a) Solid line (left y-axis): the differential resistance $R_{37,18}$ corresponding to the thermopower of the parallelogram structure at a fixed $I = +8.6$ $\mu A$ as a function of magnetic field $B$. Dotted line (right y-axis): the magnetoresistance of the parallelogram structure. (b) Solid line (left y-axis): the differential resistance $R_{73,64}$ corresponding to the thermopower of the house structure at a fixed $I = +2.1$ $\mu A$ as a function of $B$. Dotted line (right y-axis): the magnetoresistance of the house structure. The estimated thermopower scale is also drawn along the center y-axis for both structures.

The differential resistance in Fig. 3 is proportional to the thermopower by Eq. (3). Therefore, most characteristics of the thermopower of the Andreev interferometers are obtained directly from the differential resistance itself without any further analysis. However, a quantitative estimate of thermopower is still of interest. To estimate the thermopower, we need to calculate $T_m$ in Eq. (3), which is obtained from the numerical solution of Nagaev's equation (Eq. (1)). Except for $\Theta_D$ and $\alpha_{ph}$, all other parameters in Nagaev's equation are experimentally determined. We used the textbook value of $\Theta_D = 170$ K [12], and the value $\alpha_{ph} = 0.415$ obtained from shot noise experiments in Au nanowires [13], which are in good agreement with Nagaev's equation. Using these parameters, for example, $T_m$ is estimated to be 357 mK at a bath temperature $T = 351$ mK in sample A for $I = 8.6$ $\mu$A. The thermopower $S_A$ corresponding to the differential resistance data in Fig. 3 obtained from this analysis is shown on a separate scale along the center y-axis in the same figure. The magnitude of the thermopower is consistent with numerical estimates ($\sim 1$ $\mu$V/K) by Claughton and Lambert [6] for a geometry similar to the house interferometer.

Figure 4 shows the thermopower $S_A - S_N$ of the house interferometer inferred from $R_{73,64}$ measured at $B = 0$ and a fixed dc current of 2.1 $\mu$A, as a function of temperature in the range 45 - 500 mK. ($S_A - S_N$ for the parallelogram, which was measured only in a $^3$He cryostat, is consistent with these data down to 0.26 K.) The magnitude of $S_A - S_N$ shows a peak of $\sim 4$ $\mu$V/K at a temperature of $\sim 0.14$ K, decreasing rapidly below this temperature. For a pure superconductor, the ratio of the thermoelectric coefficients is predicted to be $\zeta_S/\zeta_N \sim G(x) \equiv 6/\pi^2 \int_x^\infty y^2/4\cosh^2 \frac{y}{2} \, dy$, where $\zeta = \sigma S$, $\sigma$ being the conductivity, $S$ the thermopower, $x = \Delta/k_B T$, and $\Delta$ the gap of the superconductor [15]. Although quantitative predictions of the temperature dependence have not been published, a very rough estimate can be obtained using this relation, assuming that a proximity coupled wire can be described as a superconductor with a gap determined by the correlation energy $E_c = \hbar D/L^2$, where $L$ is the length of the wire [1]. Using $L \sim 1.3$ $\mu$m, which corresponds to the distance between the node on the house interferometer and the heater wire, where the largest contribution to the thermoelectric signal might be expected, we obtain the dotted curve shown in Fig. 4. The temperature at which the minimum is predicted does not agree with our experimental data, but the overall shape of the curve is suggestively similar. We stress, however, that this idea is very speculative; a more rigorous quantitative calculation, taking into account the geometry of the device, is needed for a detailed comparison to the experiment.

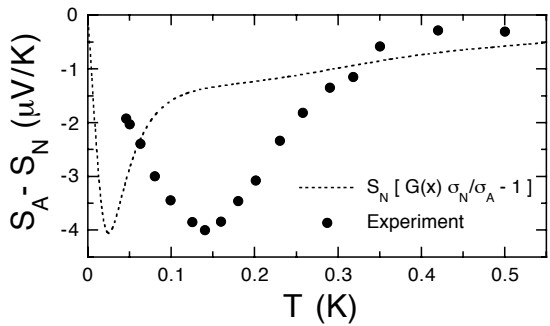

**Fig. 4.** Closed circles: thermopower $S_A - S_N$ of the house interferometer at $B = 0$ as a function of $T$, obtained from the differential resistance $R_{73,64}$ as described in the text. Dotted line: prediction for $S_A - S_N$ with $\Delta = E_c$ (as discussed in the text), where the relation $S_N \propto T$ [12] is used and $\sigma_N/\sigma_A(T)$ is obtained from experiments.

In conclusion, we have observed the phase dependence of the thermopower of an Andreev interferometer, which oscillates as a function of magnetic field with a fundamental period corresponding to a flux $\Phi_0$. This observation may expand the classical notion of thermopower into the quantum regime where the electron phase plays an important role.

We thank C. Lambert, C. Van Haesendonck, G. Neuttiens, and C. Strunk for invaluable discussions. This work was supported by the National Science Foundation through DMR-9357506, and by the David and Lucile Packard Foundation.